\documentclass[10pt]{article}
\usepackage{graphicx}

\def\Title#1{\begin{center} {\Large #1 } \end{center}}
\def\Author#1{\begin{center}{ \sc #1} \end{center}}
\def\Address#1{\begin{center}{ \it #1} \end{center}}

\newcommand\pubblock{\rightline{\begin{tabular}{l} Proceedings of the Fifth Annual LHCP\\ 
         \pubdate  \end{tabular}}}

\newenvironment{Abstract}{\begin{quotation} \begin{center} 
             \large ABSTRACT \end{center}\bigskip 
      \begin{center}\begin{large}}{\end{large}\end{center} \end{quotation}}

\newenvironment{Presented}{\begin{quotation} \begin{center} 
             PRESENTED AT\end{center}\bigskip 
      \begin{center}\begin{large}}{\end{large}\end{center} \end{quotation}}

\def\Acknowledgements{\bigskip  \bigskip \begin{center} \begin{large}
             \bf ACKNOWLEDGEMENTS \end{large}\end{center}}

\def\beq{\begin{equation}}
\def\eeq#1{\label{#1}\end{equation}}
\def\eeqn{\end{equation}}

\def\beqa{\begin{eqnarray}}
\def\eeqa#1{\label{#1}\end{eqnarray}}
\def\eeqan{\end{eqnarray}}

\let\bar=\overbar

\def\Dslash{\not{\hbox{\kern-4pt $D$}}}
\def\dslash{\not{\hbox{\kern-2pt $\del$}}}

\def\msb{{\bar{\ssstyle M \kern -1pt S}}}

\usepackage{color}

\newcommand\pubdate{\today}

\def\affiliation{
On behalf of the ALICE Collaboration, \\
Universita e INFN, Bologna, Italy }

\begin{document}

\large
\begin{titlepage}
\pubblock

\vfill
\Title{  EXHIBITING THE ALICE EXPERIMENT  }
\vfill

\Author{ DESPINA HATZIFOTIADOU  }
\Address{\affiliation}
\vfill
\begin{Abstract}

Among the many  outreach and communication tools available in our digital era, traditional tools such as exhibitions still hold an important place.  The ALICE collaboration is setting up a new exhibition at the experiment's site, as part of the ALICE Visitor Centre.  Its goal is to communicate to visitors the physics and the tools and methods used by ALICE. It combines modern technology such as video mapping with real detector items, aiming to fascinate the visitors and give them an immersive experience of a high energy physics experiment. The development process, the messages to be delivered and the choices for the contents and the way of exhibiting them are discussed; and the final design and present status of the project are presented.

\end{Abstract}
\vfill

\begin{Presented}
The Fifth Annual Conference\\
 on Large Hadron Collider Physics \\
Shanghai Jiao Tong University, Shanghai, China\\ 
May 15-20, 2017
\end{Presented}
\vfill
\end{titlepage}
\def\thefootnote{\fnsymbol{footnote}}
\setcounter{footnote}{0}
%

\normalsize 


\section{Introduction}

A variety of tools and methods are used by the ALICE collaboration for science dissemination and communication.  ALICE members give public talks and participate in events such as Open Days, European Researchers' Night and Science Fairs.  Particle physics masterclasses are held in many ALICE institutes, addressing high school students and aiming to stimulate their interest in science.  The various communication platforms available on the web (Facebook, Twitter) \cite{fb},\cite{tw}, are widely used.  Among these, visits hold a special place because they allow the visitors to see the tools of scientific research and come in contact with scientists, thus offering them a better insight into the world of science.

The ALICE experiment attracts many visitors, the highlights of the visits being the underground cavern, for which there is a big demand.  During the first long shutdown of LHC in the period 2013-2015 (LS1), more than 15 000 persons visited the ALICE cavern; during CERN Open Days in September 2013, 3500 visitors descended in two days.  Of course underground visits are only possible during limited periods, such as end year technical stops of the LHC or long shutdown periods.  Thus a surface visitor centre is essential, both for the periods when there is no access underground and also to complement the visit of the cavern.

\section{The ALICE Visitor Centre}

\begin{figure}[hbtp]
\centering
\includegraphics[height=1.8in]{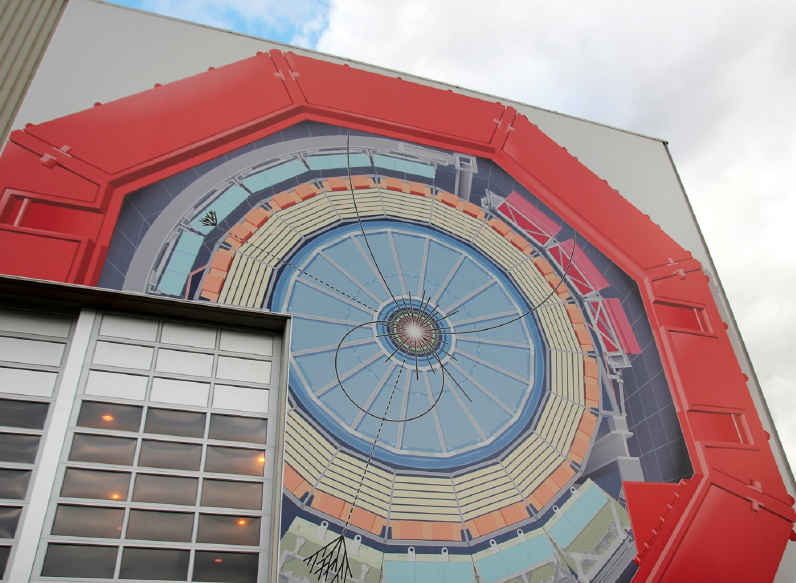}
\caption{ Full scale ALICE painting outside hall 2285 at Point 2.}
\label{fig:bigpos}
\end{figure}

 The current ALICE Visitor Centre comprises a number of items that can be seen during a surface visit at Point 2.  Outside hall 2285 there is a full scale painting of the ALICE cross section (Figure \ref{fig:bigpos}); this gives a first idea of the dimensions of the experiment and helps to introduce the different components of the detector and their role.
\begin{figure}[hbtp]
\centering
\includegraphics[height=2.in]{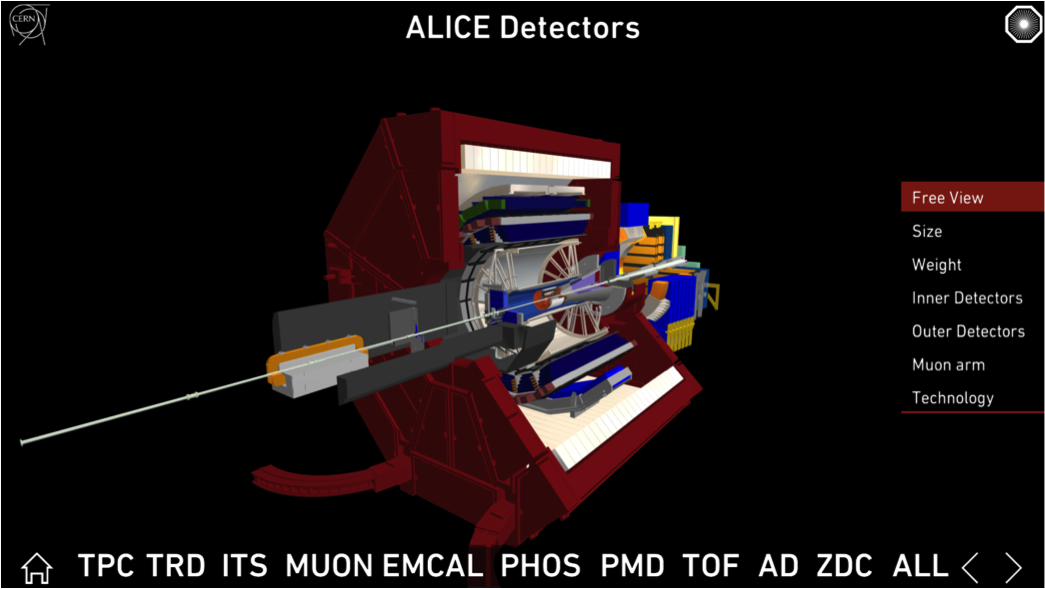}
\caption{ The ALICE 3-D model on the interactive window at the ARC.}
\label{fig:int-win}
\end{figure}
  Entering the hall, the visitor can get a glimpse of the Alice Run Control Centre (ARC) through the windows and see the shifters at work; the guide can discuss the LHC operation and the data taking.  One of the ARC windows is an interactive window; by pressing a button it becomes opaque and serves as a touch screen; there the guide can describe the various detector subsystems (Figure \ref{fig:int-win}) and give information about the ALICE member institutes.
 Walking past the ARC, the visitor can see the point of access to the cavern (PAD and MAD), the shaft, and the counting rooms housing the DAQ and DCS computers, and visit the new ALICE exhibition.

\section{The ALICE exhibition}

The ALICE exhibition was first created at Point 2 on the occasion of the Open Days for CERN's 50th anniversary.  After serving its purpose for many years, it needed to be rejuvenated, using modern technologies.  One of the aims was to integrate ALICE in the official sites of the CERN visits service, so it had to comply with their requirements.  Its location, close to the experiment and its control room, offers to visitors the unique experience to see and feel a real experimental environment.

 Our goal was to come up with a concept satisfying both the requirements of general public visits and the specific needs of the ALICE collaborators.  Our objective was to present to visitors the physics of relativistic heavy ion collisions and its relevance for understanding the early universe as well as the various tools and methods used by the ALICE experiment. 

The development of the concept for the new exhibition was done in many stages.   A working group consisting of ALICE members met, brainstormed and defined the main topics to be addressed: what ALICE does (the physics), how it does it (the detectors) and what it has observed (the results).  It was agreed that the best way to introduce these concepts would be by showing an animation; so the visit will start with viewing a film; then a number of exhibits, mainly detector items, can be shown and explained by the guide. 

The ALICE technical coordination embraced the project and brought in new ideas.  We had fruitful discussions with the CERN Education, Communications and  Outreach group.  The various CERN exhibitions and visit points were also a useful source of inspiration.  In the end we decided to seek the help of professionals, and opted for the Spanish company Indissoluble\cite{indi}, specializing in exhibitions, who has done the new CERN microcosm.
\begin{figure}[hbtp]
\centering
\includegraphics[height=2in]{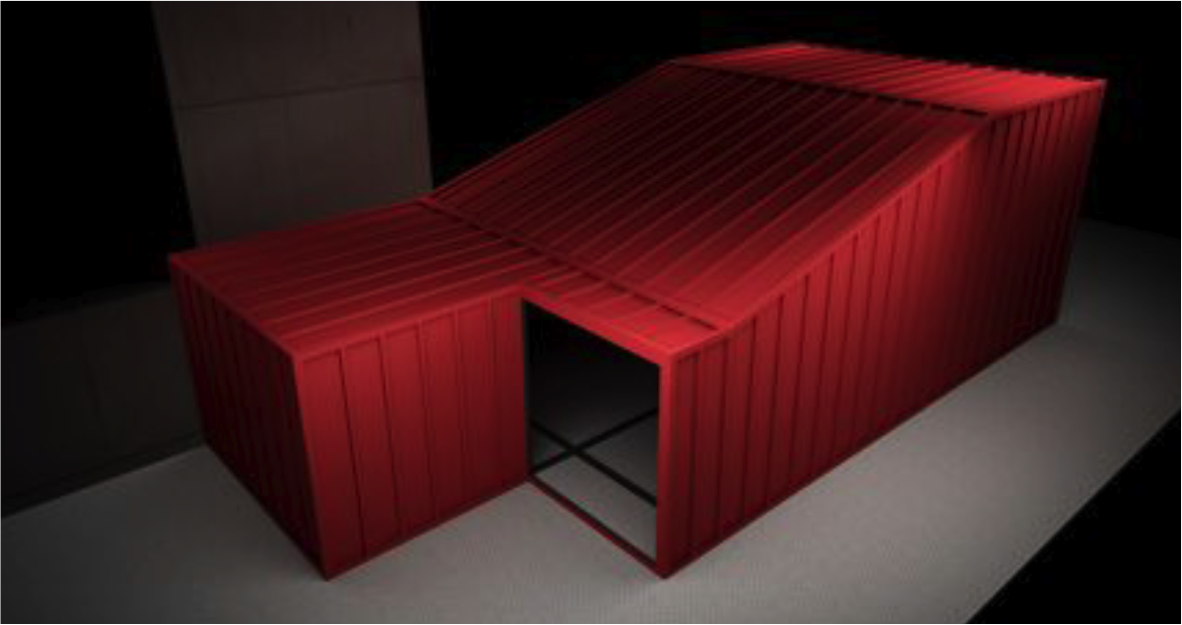}
\caption{ The structure housing the exhibition.}
\label{fig:redstructure}
\end{figure}

\begin{figure}[hbtp]
\centering
\includegraphics[height=2in]{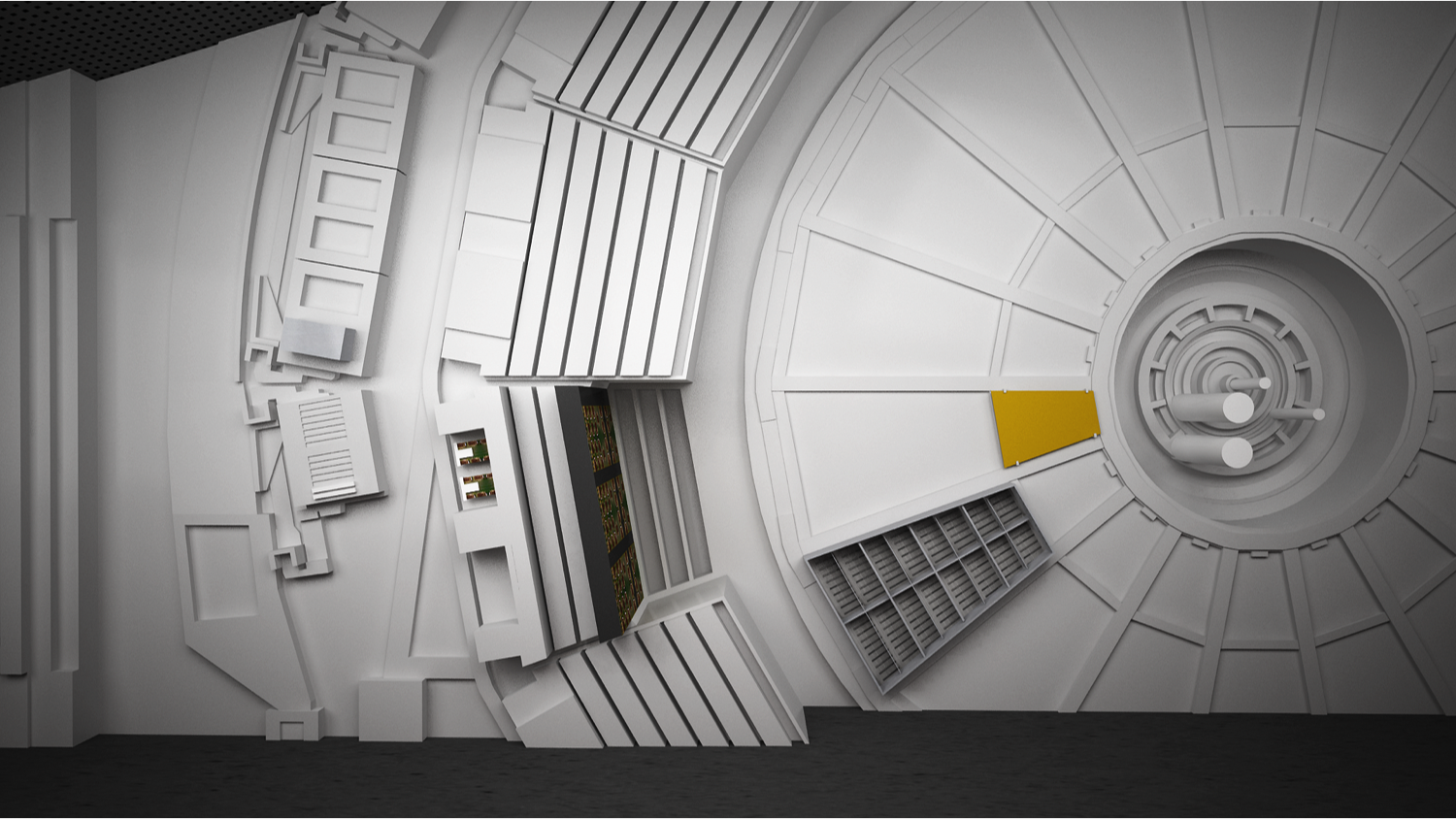}
\caption{ The ALICE cross-section mockup with embedded detectors.}
\label{fig:fig-mockup}
\end{figure}

\begin{figure}[hbtp]
\centering
\includegraphics[height=1.8in]{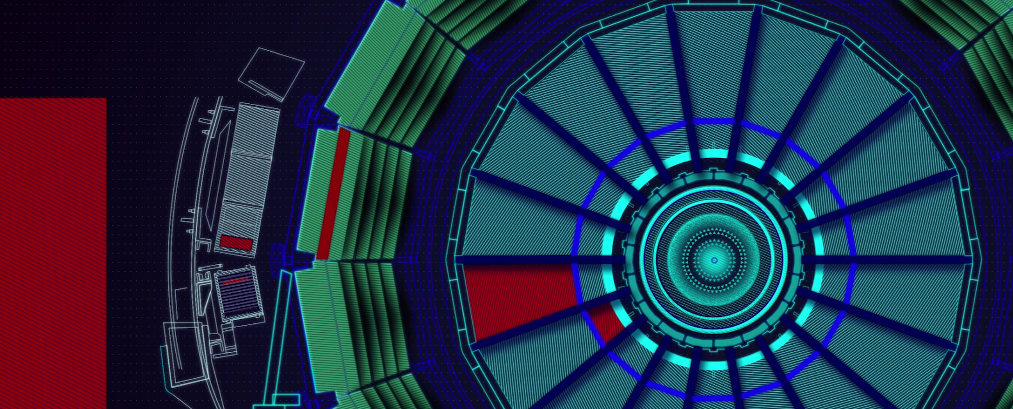}
\caption{ The projection mapping on the ALICE cross-section mockup.}
\label{fig:fig-map}
\end{figure}

\begin{figure}[hbtp]
\centering
\includegraphics[height=3.1in]{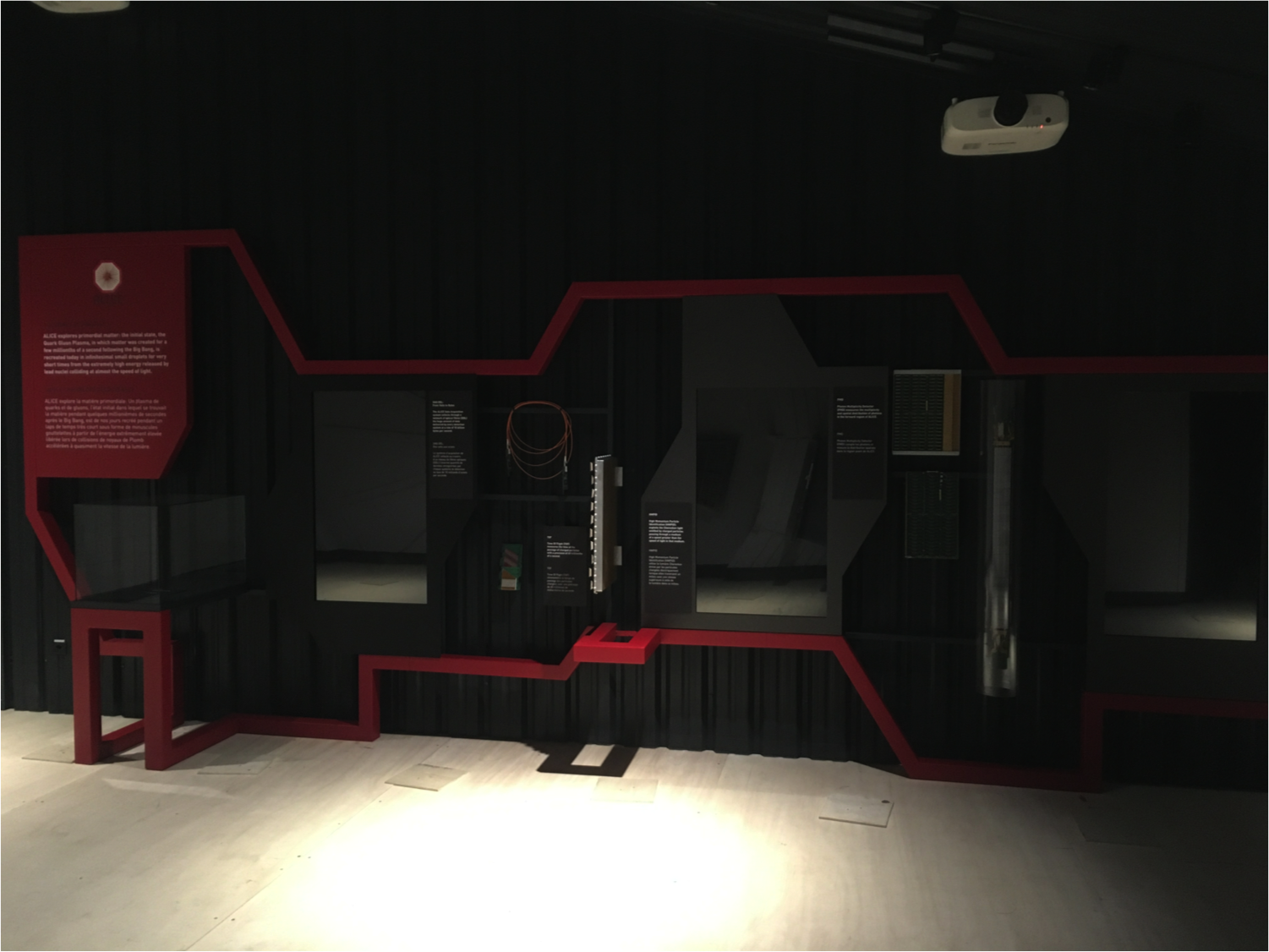}
\caption{ The showcases inside the exhibition.}
\label{fig:fig-vit}
\end{figure}

The site of the existing exhibition was an open space, approximately 100 m$^2$, sitting next to the shaft, on top of the counting rooms.  The new exhibition is installed at the same spot, inside a closed structure (Figure \ref{fig:redstructure}), which ensures acoustic insulation from the noisy environment at Point 2 and allows controlled illumination of the exhibits.  One of the long walls is covered with a wooden, real size mockup of the ALICE cross section, with dimensions 11 m x  5 m.  Detector items are embedded in the mockup in places corresponding to their real position inside the experiment (Figure \ref{fig:fig-mockup}).  Video projection mapping brings the ALICE detectors to life (Figure \ref{fig:fig-map}).
Additional exhibits are installed inside showcases, on the wall opposite the mockup (Figure \ref{fig:fig-vit}); three touch screens display information about the exhibits and help the guide explain their operation.  The exhibits present the various functions of particle detectors (tracking, calorimetry, particle identification) and cover a broad spectrum of technologies (gaseous detectors, scintillators, silicon detectors, etc).

\begin{figure}[hbtp]
\centering
\includegraphics[height=2.1in]{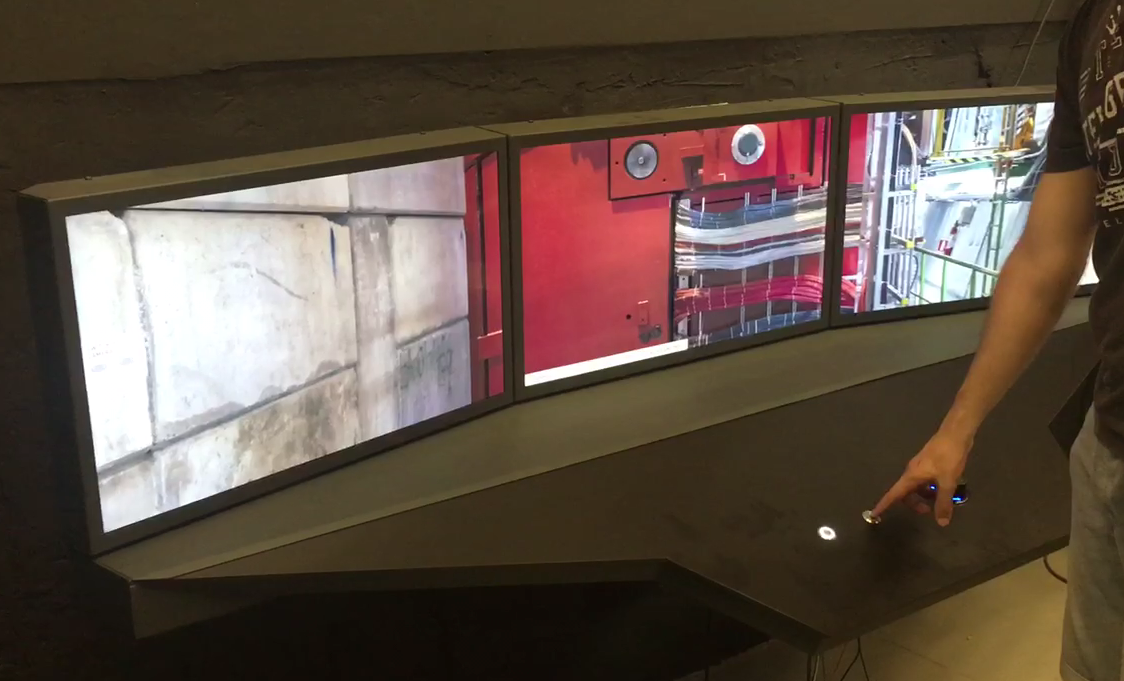}
\caption{ View of the ALICE detector through a computer-controlled camera.}
\label{fig:fig-per}
\end{figure}

Another attraction inside the exhibition is a simulation of a periscope, offering to the visitors a direct view of the cavern.  Two computer-controlled cameras are installed underground on the two sides of ALICE: one facing the magnet doors, on the so-called low-beta platform; and another one on the side of the muon spectrometer.  They  are connected with three screens; the visitor can select one of the two cameras, rotate it and zoom; additional high-resolution photographs of what the camera sees will offer details to the curious (Figure \ref{fig:fig-per}).

\begin{figure}[hbtp]
\centering
\includegraphics[height=2.2in]{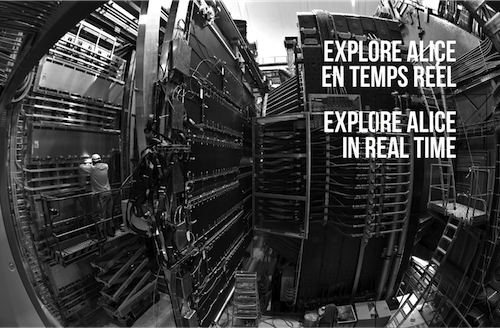}
\caption{ The photograph of the muon spectrometer.}
\label{fig:muon-sp}
\end{figure}

The forward muon spectrometer of ALICE is shown as a large photograph (Figure \ref{fig:muon-sp}).

The control of the lights, the video mapping projection and the interactive part (touch screens and screens connected to the underground cameras) is done by an iPad.  This communicates with a central server and a number of PCs for the projection and the control of the cameras in the cavern.
 



\section{The introductory film}

The introductory film, seven minutes long, is an animation with narration in english and french; more languages will follow.  The storyboard, developed over many discussions and finalized after many iterations, consists of seven chapters: introduction, about matter, Big Bang cosmology, how the Quark Gluon Plasma (QGP) 
is produced; how we observe the debris of the QGP;  how the experiment works - from data collection to publication; collaboration, results and conclusion.  This animation was done by Indissoluble, with the invaluable help of the CERN ECO group and Media Lab.

\begin{figure}[hbtp]
\centering
\includegraphics[height=2in]{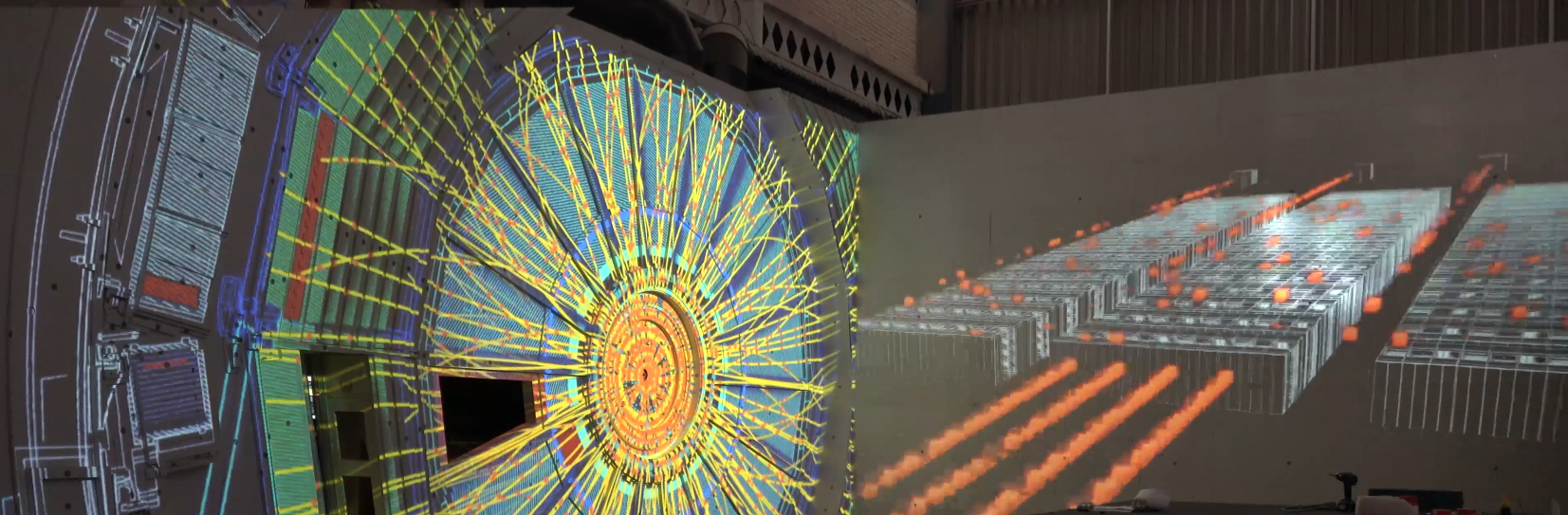}
\caption{ Projection on the mockup and the wall perpendicular to it.}
\label{fig:pro}
\end{figure}

The projection is done with three projectors on a wall perpendicular to the mockup and on the mockup itself (Figure \ref{fig:pro}).

\section{Current status and the future}

At the time of writing (October 2017) the installation is finished while some fine adjustments are done for the video mapping; the content for the touch screens is also under development.   When everything is ready and tested with some pilot visits, the exhibition will be opened to the public.  The visitors will be accompanied at all times, so guides will need to be trained, both for practical issues and to become familiar with the ALICE physics and detectors. 

\section{Conclusions}

We hope that the ALICE Visitor Centre and in particular the new exhibition will increase the visibility of the ALICE experiment and will also help bring the public closer to the fascinating physics of heavy ion collisions, which is an integral part of CERN's rich research programme. 

\Acknowledgements
We are  grateful to Rolf Landua of CERN IR-ECO for following this project all along its development, for numerous discussions, and for his invaluable contribution to the making of the introductory film.  We have greatly profitted from his experience.  We also thank Emma Sanders for fruitful discussions, and Daniel Dominguez for his contribution to the animation.

\end{document}